\documentclass[aps
,%
prl,%
twocolumn%
]{revtex4}

\usepackage[dvips]{graphicx}
\def\gsim{\buildrel {\textstyle >}\over {_\sim}}
\def\lsim{\buildrel {\textstyle <}\over {_\sim}}

\begin{document}
\title{Simulation study of spatio-temporal correlations of earthquakes 
as a stick-slip frictional instability
}
\author{Takahiro Mori and Hikaru Kawamura}
\affiliation{Department of Earth and Space Science, Faculty of Science,
Osaka University, Toyonaka 560-0043,
Japan}
\date{\today}
\begin{abstract}
Spatio-temporal correlations of earthquakes  
are studied numerically 
on the basis of the one-dimensional spring-block (Burridge-Knopoff) model.
As large events approach, the frequency of smaller 
events gradually increases, 
while, just before the mainshock, it is dramatically
suppressed in a close vicinity of the epicenter of the upcoming mainshock,
a phenomenon closely resembling the ``Mogi doughnut''.
\end{abstract}
\maketitle

Earthquake is a stick-slip frictional instability of a fault
driven by steady motions of tectonic plates\cite{ScholzRev,ScholzBook}.
While earthquakes are obviously  complex phenomena, 
certain empirical laws have been known concerning their statistical properties,
{\it e.g.\/}, the Gutenberg-Richter (GR) law for the magnitude distribution of
earthquakes, or
the Omori law for the time evolution of the frequency of 
aftershocks\cite{ScholzBook}. These laws are basically
of statistical nature, becoming eminent only after analyzing large number of
events. 

Since earthquakes could be regarded as a stick-slip frictional 
instability of
a pre-existing fault, the statistical properties of earthquakes 
should be governed  by the physical law of 
rock friction\cite{ScholzRev,ScholzBook}. 
One might naturally ask: How the statistical properties of earthquakes 
depend on the material properties characterizing earthquake faults, 
{\it e.g.\/},
the elastic properties of the crust or the frictional properties of 
the fault, {\it etc\/}. 

Some time ago, Carlson, Langer and collaborators
performed a pioneering study of 
the statistical properties of earthquakes \cite{CL,CLST,Carlson}, 
based on the spring-block model (Burridge-Knopoff model) \cite{BK}.
These authors paid particular attention to
the magnitude distribution of earthquake events, and 
examined its dependence on the friction parameter characterizing the 
nonlinear stick-slip dynamics of the model. It was observed that,
while smaller events persistently obeyed the GR law, {\it i.e.\/},
staying critical or near-critical, 
larger events exhibited a significant 
deviation from the GR law, being off-critical 
or ``characteristic''.\cite{CL,CLST,Carlson,Schmittbuhl}  
Shaw, Carlson and Langer studied the same model by
examining the spatio-temporal patterns
of seismic events preceding large events, observing that
the seismic activity accelerates as the large event approaches\cite{Shaw}.
Recently, many numerical works have been made for the 
cellular-automaton versions of the model which were introduced to mimic the
original spring-block model \cite{OFC,Hergarten,Hainzl}.

In the present Letter, we wish to further investigate
the spatio-temporal correlations of earthquakes
by performing extensive numerical simulations of the 
Burridge-Knopoff (BK) model \cite{BK}.
Our main  goal is to clarify how the statistical properties
of earthquakes depend on the friction parameter characterizing the 
nonlinear stick-slip dynamics.
%
We simulate here the one-dimensional (1D) version of the 
BK model. It consists of a 1D array of $N$ identical blocks, 
which are mutually connected with the two neighboring 
blocks via the 
elastic springs of the 
elastic constant $k_c$, and are also connected to 
the moving plate
via the springs of the elastic constant $k_p$.
All blocks are subject to the 
friction force, which is the only source of the nonlinearity in the
model. The time $t$ is made dimensionless here, being measured in units of
the characteristic frequency $\omega =\sqrt{k_p/m}$ where $m$ is the
mass of a block. Then, the equation of motion for the $i$-th block can be
written in the dimensionless form as
\begin{equation}
\ddot u_i=\nu t-u_i+l^2(u_{i+1}-2u_i+u_{i-1})-\phi (\dot u_i),
\end{equation}
where $u_i$ is the dimensionless displacement of the 
$i$-th block, 
$l \equiv \sqrt{k_c/k_p}$ is the dimensionless stiffness parameter, 
$\nu $ is the dimensionless loading rate 
representing the speed of the plate, and  
$\phi$ is the dimensionless friction force.
In order for the model to exhibit a 
dynamical instability corresponding to an earthquake, it is essential 
that the friction force $\phi$ possesses a frictional {\it weakening\/}
property, {\it i.e.\/},
the friction should become weaker as the block slides.
Here, as the form of the frictional force, 
we assume the form used in Ref.\cite{CLST}, which represents the
velocity-weakening friction force;
\begin{equation}
\phi(\dot u) = \left\{ 
             \begin{array}{ll} 
             (-\infty, 1],  & \ \ \ \ {\rm for}\ \  \dot u_i\leq 0, \\ 
              \frac{1-\sigma}{1+2\alpha \dot u_i/(1-\sigma )}, &
             \ \ \ \ {\rm for}\ \  \dot u_i>0, 
             \end{array}
\right.
\end{equation}
where its maximum value corresponding the static friction
is normalized to unity. This normalization 
condition $\phi(\dot u=0)=1$ has been utilized to set the length unit.
The back-slip is inhibited by imposing an
infinitely large friction for $\dot u_i<0$, {\it i.e.\/}, 
$\phi(\dot u<0)=-\infty $. 

The friction force is characterized by the two parameters, $\sigma$ and 
$\alpha$. The former, $\sigma$, 
represents an instanteous drop of the friction force
at the onset of the slip, while the latter, $\alpha$, 
represents the rate of the friction force getting weaker
on increasing the sliding velocity. Following Ref.\cite{CLST}, 
we assume the loading rate $\nu$ to be infinitesimally small, and put 
$\nu=0$ during an earthquake event. Taking this limit ensures that the interval
time during successive
earthquake events can be measured in units of $\nu^{-1}$ 
irrespective of 
particular values of $\nu$ \cite{CLST}.  

Although we studied the properties of the model with varying all the parameters
$\alpha$, $\sigma$ and $l$,  we explicitly
show here the $\alpha$-dependence only, since the parameter $\alpha$ 
turns out to affect the result most significantly. 
We mainly study the cases $\alpha =1,2$ and 3: This choice has been 
made because (i) Ref.\cite{CL} suggested that 
the smaller value of $\alpha <1$ tended to cause 
a creeping-like behavior, 
and (ii) our preliminary study for larger $\alpha$
($\alpha =5, 10$) indicated that the further increase of $\alpha >3$ did not 
change the result qualitatively.
Below, we fix $l=3$ and $\sigma=0.01$ 
unless otherwise stated. 
In order to eliminate the possible finite-size effects, the total
number of blocks are taken to be large, typically $N=800$, with
periodic boundary conditions. In the case of $l=3$ and $\sigma=0.01$,
even the largest event in our simulations
involves the number of blocks less than $N=800$.
Total number of $10^7$  events
are generated in each run.

In Fig.1 we show the magnitude
distribution $R(\mu)$ 
of earthquake events for several values of the parameter $\alpha$, 
where $R(\mu){\rm d}\mu$ represents the rate of events with their
magnitudes in the range [$\mu, \mu +{\rm d}\mu$].
The magnitude of an event, $\mu$, 
is defined as the logarithm of the moment $M_0$, {\it i.e.\/}, 
$\mu=\ln M_0$ with $M_0=\sum_i \Delta u_i$,
where $\Delta u_i$ is the 
total displacement of the $i$-th block during a given 
event and the sum is taken over all blocks involved in the event\cite{CLST}.

As can be seen from Fig.1,
the data  for  $\alpha =1$ lie on a straight line fairly well, apparently
satisfying the GR law. The values of the exponent $B$ describing 
the power-law behavior, 
$\propto 10^{-B}$, is estimated to be $B\simeq 0.50$.
The data for larger $\alpha$, {\it i.e.\/}, for $\alpha =2$ and 3, 
deviate
from the GR law at larger magnitudes, exhibiting a clear peak structure, 
while the power-law feature still remains for smaller magnitudes.
These features of the magnitude distribution are consistent with the
earlier observation of Carlson and Langer \cite{CL,CLST}.
The observed peak structure 
gives us a criterion to distinguish large and small events. Below,
we regard events with their magnitudes $\mu$ greater than $\mu_c=3$
as large events, $\mu_c=3$ being 
close to the peak position of the magnitude distribution of Fig.1. 
In an earthquake with  $\mu=3$,
the mean number of moving blocks are about 76 ($\alpha =1$) and 60 
($\alpha =2,3$).

\begin{figure}[ht]
\begin{center}
\includegraphics[scale=1.4]{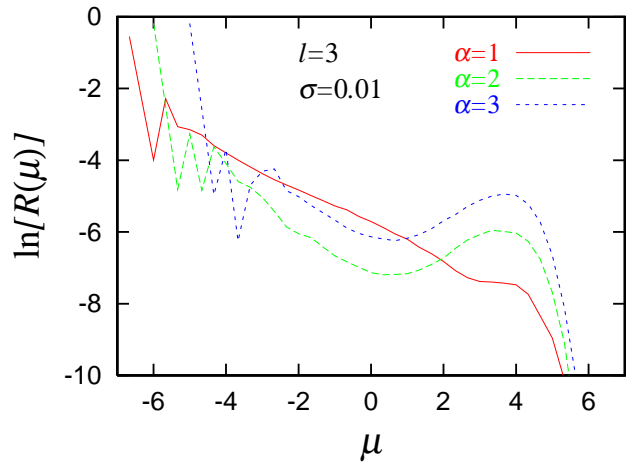}
\end{center}
\caption{
Magnitude distribution of earthquake events for various values of $\alpha$.
}
\end{figure}

One question of general interest 
is how large earthquakes repeat in time, 
do they occur near periodically or irregularly?
One may ask this question either locally, {\it i.e.\/}, for a given finite 
area on the fault,
or globally,  {\it i.e.\/}, for an entire fault system.
In Fig.2, we show the distribution of the
recurrence time $T$ of large earthquakes with $\mu _c=3$, 
measured either locally (a), or
globally (b). In the insets, the same data including the tail part 
are re-plotted on a 
semi-logarithmic scale.
In defining the recurrence time locally, 
the subsequent large event is counted when a large
event occurs with its epicenter in the region within
30 blocks from the epicenter of the previous large event.
The mean recurrence time $\bar T$ is then estimated to be
$\bar T\nu =1.47$, 1.12, and 1.13 for $\alpha=1$, 2 and 3,  respectively.

%
\begin{figure}[ht]
\begin{center}
\includegraphics[scale=0.65]{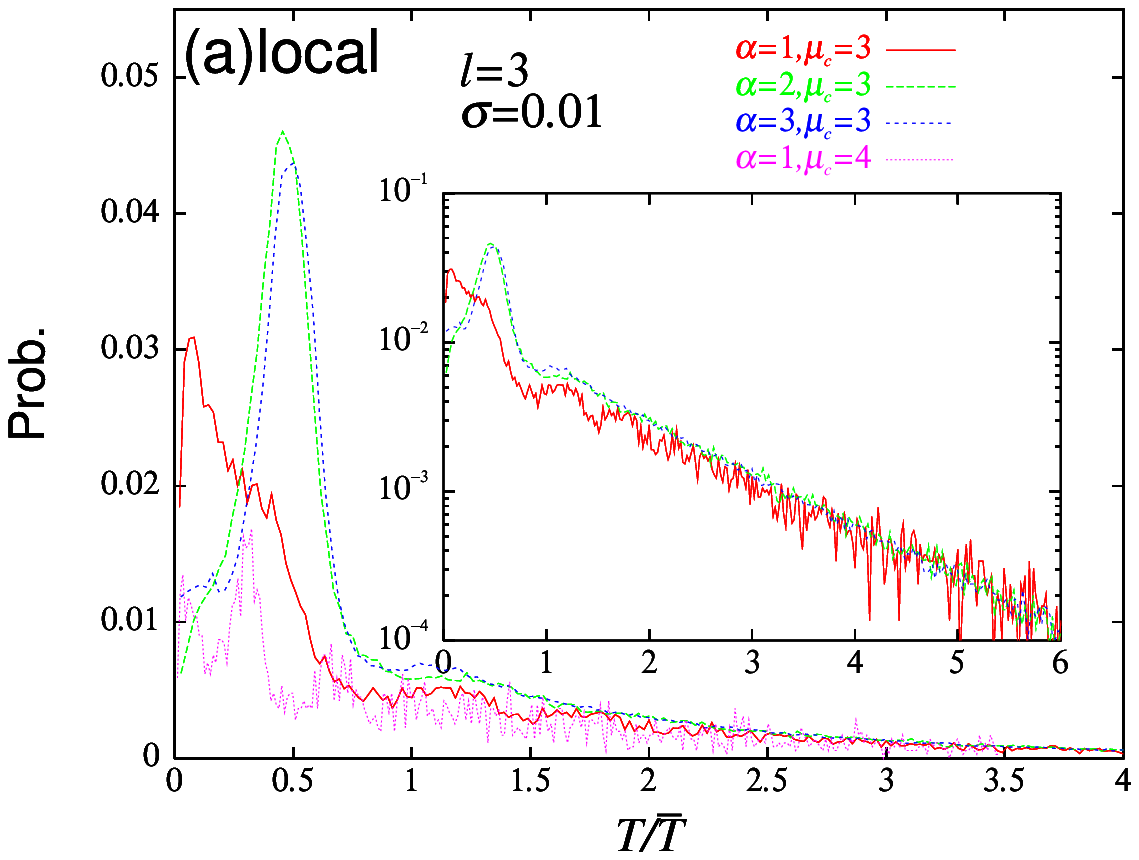}
\includegraphics[scale=0.65]{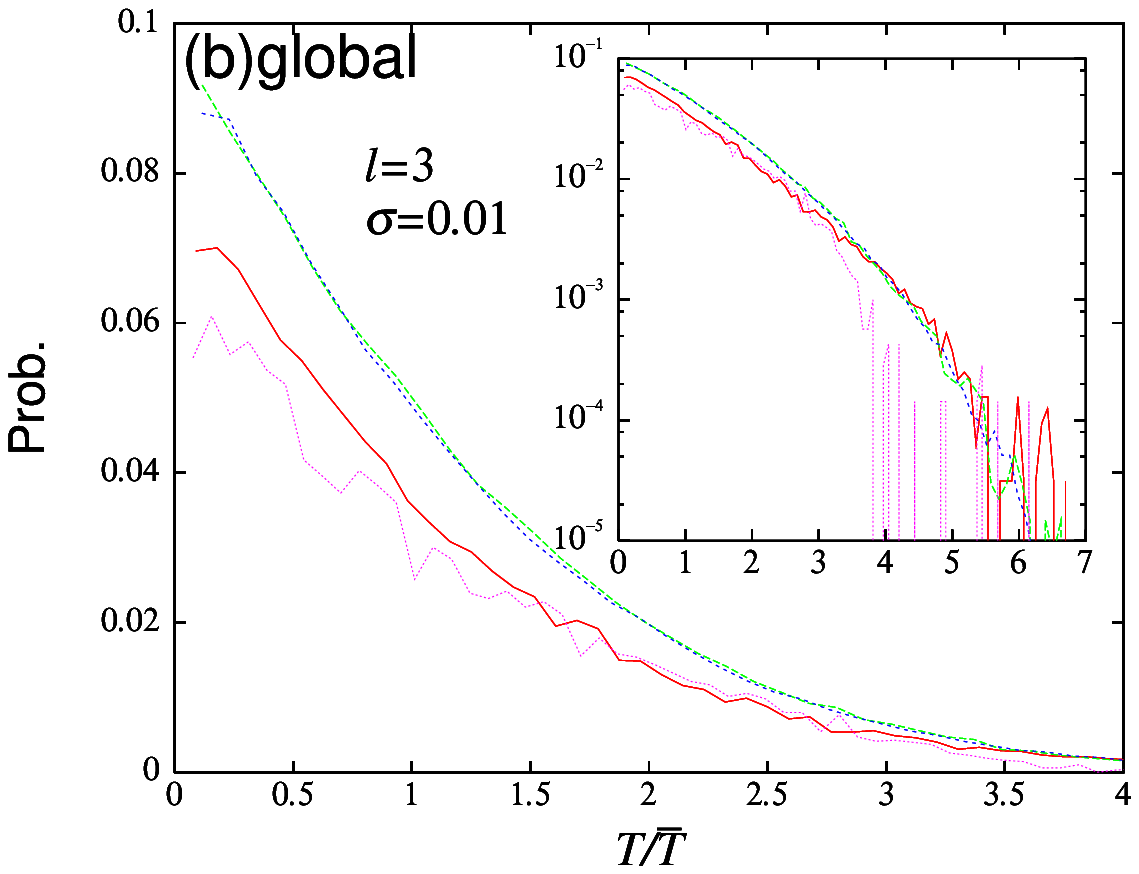}
\end{center}
\caption{
The recurrence-time distribution of large events with their magnitudes
greater than $\mu_c=3$ or 4 for various values of $\alpha$, the local
one (a) and the global one (b). The recurrence time $T$ is normalized by its mean $\bar T$. The total number of blocks is $N=800$. Even the largest event involves the number of blocks less than $N=800$. The insets represent the semi-logarithmic plots including the tail part of the distribution.
}
\end{figure}

The local recurrence-time
distribution shown in Fig.2(a) has the following two noticeable features. 
(i) The tail of the distribution  is exponential at longer $T\gsim \bar T$
for all values of $\alpha$. 
Such an exponential tail of the distribution has also been
reported for real faults\cite{Corral}.
(ii) The form of the distribution at shorter $T\lsim \bar T$ 
is non-exponential,
and largely differs between
for $\alpha =1$ and for $\alpha =2$ and 3. 
For $\alpha=2$ and 3, the distribution
has an eminent  peak at around
$\bar T\nu \simeq 0.5$, not far from the mean  
recurrence time.
Here we regard such an appearance of a characteristic recurrence time as a
signature of the near-periodic recurrence of 
large events. Indeed, such a near-periodic recurrence of large
events was reported for several real faults \cite{ScholzBook,NB}.

For $\alpha =1$, by contrast,
the peak located close to the mean $\bar T$ is hardly discernible. 
Instead, the distribution has a pronounced peak at a 
shorter time $\bar T\nu \simeq 0.10$, just after the previous large event.
In other words, large events for $\alpha=1$ tend to occur as 
 ``twins''. This has also been confirmed by our analysis of the time record
of large events.
Closer examination of individual event has revealed that 
a large event for $\alpha=1$
often occurs as a ``unilateral earthquake'' where the rupture propagates
only in one direction, hardly propagating 
in the other direction. In other words,
for $\alpha =1$, the epicenter of large events tend to be located
near the edge of the rupture zone. 
This can be confirmed more
quantitatively by calculating the ``eccentricity'' 
of the epicenter $\epsilon =R^*/\bar R$, which is defined
by the ratio of the mean distance between the epicenter and the center of mass
of the rupture zone, $R^*$,
to the mean radius of the rupture zone, $\bar R$. 
Then, we have found $\epsilon = 0.88$, 0.52 and 0.53
for the cases $\alpha =1$, 2 and 3, respectively.
The occurrence of unilateral earthquakes for smaller $\alpha$
may be understandable if one notes that the relative
weakness of the stick-slip instability prevents the initiated rupture
from propagating far into both directions. 
When a large earthquake occurs in the form of  
such a unilateral earthquake, further
loading due to the plate motion tends to trigger the subsequent
large event in the opposite direction, causing a twin-like event. 
This naturally explains the small-$T$ peak observed in Fig.2(a) for $\alpha=1$.

We note, however, that even in the case of $\alpha =1$ 
the periodic character of events
becomes appreciable when one looks at very large events. In Fig.2(a), the
recurrence-time distribution for very large events, characterized by 
still larger magnitude
threshold $\mu _c=4$, is shown  for the case of $\alpha =1$. 
Interestingly, 
the distribution in this case has {\it two\/} 
distinct peaks, one corresponds to 
the twin-like event and the other corresponds to the near-periodic event.
Hence, even in the case of $\alpha=1$ where the critical features
are apparently dominant, features of characteristic earthquake becomes 
increasingly eminent when 
one looks at very large events.

%
%
%
%

The {\it global\/} recurrence-time distribution, 
{\it i.e.\/}, the one for an entire
fault system with $N=800$, takes a
different form from the local one, as can be seen from Fig.2(b). 
The peak structure seen in the local distribution no longer 
exists here. 
Furthermore, 
the form of the distribution tail at larger $T$
is no longer a simple exponential, faster than exponential:
See a curvature of the data in the inset of Fig.2(b). 

Carlson  \cite{Carlson} and Schmittbuhl {\it et al\/} \cite{Schmittbuhl}
reported a periodic behavior of large events 
by studying the {\it global\/} distribution function of the model 
of smaller size $N=100$ with
$l =10$. We have found that, 
for such a  large value of $l$, even an $N=800$ system behaves 
almost as a rigid body and exhibits a near-periodic behavior, where the
larger events often penetrates the entire $N=800$ system.
We note that, generally speaking, whether the recurrence-time
distribution exhibits a periodic peak depends on the length scale
of measurements as well as on the values of the parameters $\alpha $
and $l$. Such scale-dependent
features of the
recurrence-time distribution of the BK model is in apparent contrast with the
scale-invariant features of the
recurrence-time distributions recently reported for some of real faults 
\cite{Bak,Corral}. 

%
\begin{figure}[ht]
\begin{center}
\includegraphics[scale=0.65]{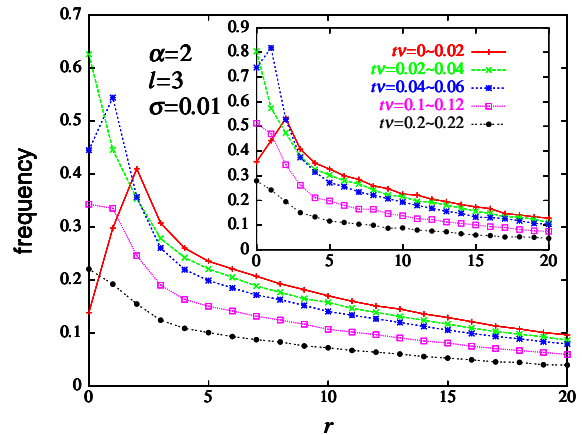}
\end{center}
\caption{
Event frequency preceding the large event with $\mu >\mu_c=3$ plotted
versus $r$,
the distance from the epicenter of the upcoming mainshock, for
$\alpha=2$ and for several time periods before the mainshock.
The inset represents the weighted event frequency with the weight of the
event size (the number of blocks): See the text for details.
}
\end{figure}
%
%

In order to ``predict'' the large event, one usually probes the
precursory phenomena associated with the large event.
Fig.3 represents the space-time correlation function 
between the large events and the preceding 
events of arbitrary size (dominated in number
by smaller events): 
It represents the conditional probability that,
provided that a large event with $\mu >\mu_c =3$ 
occurs at a time $t_0$ and at a spatial
point $r_0$,
an event of arbitrary size occurs at a time $t_0-t$ and at a spatial 
point $r_0\pm r$. The calculated correlation functions  
for the case of $\alpha =2$
are shown as a function of
$r$ for several time periods before the mainshock.

As can be seen from Fig.3, preceding the large event, 
there is a clear tendency
of the frequency of smaller events to be enhanced at and around 
the epicenter of the upcoming mainshock \cite{Shaw}. 
For small enough $t$, such a cluster of smaller events correlated
with the large event may be regarded as foreshocks. 
An interesting feature revealed here
is that, as the mainshock becomes imminent, the frequency of
smaller events is {\it suppressed\/} in a close vicinity of the epicenter of the
upcoming mainshock, though it continues to be enhanced in the surroundings.
For real earthquake faults, such a quiescence phenomenon 
has been discussed as the ``Mogi doughnut''\cite{Mogi,ScholzBook}. 

One may wonder if
our way of measuring the seismicity by 
simply counting the number of events may over-weigh 
the contribution of the 
single-block events  
which are by far the most frequent events.
In order to examine this point, we show in the inset of Fig.3
the weighted correlation function in which the frequency of small events
is counted with the weight proportional to its size 
(the number of blocks involved in that event). To eliminate the 
contribution of the large event, we count here only the small 
events with its size less than 10 blocks. 
As can be seen from the inset, essentially the
same quiescence phenomena as shown in the main panel have been observed,
suggesting that the quiescence observed here is a robust property of the
model.
Simulations done with varying the $\alpha$-values have revealed that
such a Mogi-doughnut quiescence tends to to be enhanced with
increasing $\alpha$.

\begin{figure}[ht]
\begin{center}
\includegraphics[scale=1.4]{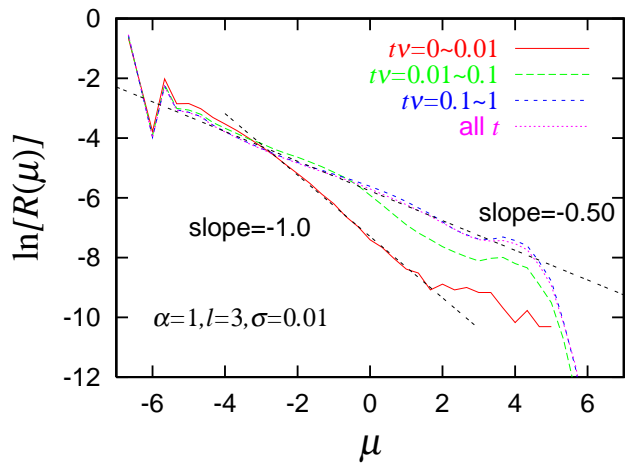}
\end{center}
\caption{
Local magnitude distribution preceding the mainshock with $\mu >\mu _c=3$,
for $\alpha=1$ and for several time periods before the mainshock.
}
\end{figure}

The present observation contrasts with the earlier work of Carlson,
who claimed that the 1D BK model did not exhibit such a Mogi-doughnut 
quiescence \cite{Carlson}.
We note that the quiescence observed here 
occurs only in a close vicinity of the 
epicenter of the mainshock, 
within one or two blocks from the epicenter, and only 
at a time close to the mainshock, $t\nu \lsim 0.02$.
Thus, 
the better statistics and the better resolution of the present simulation
have been essential to uncover this effect.


%

As an other signature of the precursory phenomena,
we show in Fig.4 the ``time-resolved''  local magnitude distribution
for the case of $\alpha =1$ for several time periods before the large
event. Only the events with their epicenters within 30 blocks from the
upcoming mainshock is counted here.
As can be seen from the figure, as the mainshock approaches, 
the form of the magnitude distribution changes significantly. In particular,
the apparent $B$-value describing the 
power-law regime tends to {\it increase\/}  as the mainshock
approaches, from the time-averaged mean value $\simeq 0.50$ to the value
$\simeq 1.0$ just before the mainshock: It is almost doubled.
Interestingly, a 
similar increase of the apparent $B$-value
preceding the mainshock was reported for real faults \cite{Smith}. 
For the case of larger $\alpha$, $\alpha =2$ and 3, 
the change of the $B$-value preceding the mainshock is still appreciable 
(the data not shown here), though in a less pronounced manner. 
While the observed change in the magnitude distribution might simply
be understood as caused by 
a supression of larger events prior to the mainshock, we
emphasize it is a real observable effect if one monitors the time change
of the local magnitude distribution.


In summary, we studied the spatio-temporal correlations of the 1D BK model of
earthquakes. Periodic feature of large events is eminent when
the friction force exhibits a strong frictional instability, whereas,
when the friction force exhibits a weak frictional instability, 
large events often occur as twin and/or unilateral events.
Preceding the mainshock, the frequency of smaller 
events is gradually
enhanced, whereas, just before the mainshock, it is dramatically
suppressed in a close vicinity of the epicenter of the upcoming mainshock
(the Mogi doughnut). Under certain conditions, preceding the mainshock, 
the apparent $B$-value of the magnitude distribution increases significantly.
These properties may be used in predicting the time and the position of 
the upcoming large event.

\end{document}